# The CONVERGE project – A systems network for ITS

**Prof. Dr. Horst Wieker**
*Hochschule für Technik und Wirtschaft des Saarlandes, Goebenstraße 40,66115, Saarbrücken, Germany*

**Manuel Fünfrocken, M.Sc.**
*Hochschule für Technik und Wirtschaft des Saarlandes, Altenkessler Straße 17/D2,66117, Saarbrücken, Germany*

**Jonas Vogt, M.Sc.**
*Hochschule für Technik und Wirtschaft des Saarlandes, Altenkessler Straße 17/D2,66117, Saarbrücken, Germany*

**Abstract**

Various German and European research projects [1,2,3,4] in the past have create several isolated solutions for intelligent transportation systems (ITS) infrastructure. As a matter of fact, a lot of those solutions are temporary. The backend architecture provided by those systems is mostly focused on only supporting test and evaluation of topics in the context of vehicular based application and communication.

There have been a couple of research projects, which have evaluated the different communication paths between ITS roadside or central stations and vehicle stations [1,3,5]. The current approach of the research community is to combine multiple ways of communication to provide the greatest benefit. Two of the most promising technologies are cellular communication, also called ETSI ITS IMT Public [6], and Car-to-Car communication (C2C), further called ETSI ITS G5 [7]. As there have already been research projects which evaluate the possibilities of one of those communication paths, few research projects have evaluated how those paths could complement each other. As the implementation and integration of new business cases should be as easy as possible, the resulting ITS services should have a standardized, open and easy to access interface to current communication paths as well as to those paths enabled by future technologies.

The structure of the Car2X Systems Network will be based on the Internet, in which different and independent participants are working equally together over standardized interfaces and protocols. However, the Car2X Systems Network needs some special solutions, as the concepts of the Internet are not entirely applicable to the requirements of the ITS world. Especially the strict security and privacy considerations of ITS environments are not or only insufficiently addressed in current state-of-the-art internet services. Furthermore, network requirements like Quality of Service (QoS) needs also to be considered. Thus the Car2X Systems Network aims to provide open, standardized, secure and interchangeable interfaces for each possible participant of the ITS environment.

The goal of CONVERGE is, to create a reference architecture for the Car2X Systems Network. Besides the basic network architecture, security and privacy concepts need to be created. Consequently, CONVERGE will create an effective, secure, scalable and marketable architecture for ITS and deliver valuable input to standardization.

**Initial situation: State-of-the-art**

ITS services, which are currently available to customers, tend to be closed solutions which tightly couple the different layers of those solutions: the service logic itself (service), the communication architecture (communication way) including an optional communication center as well as the application logic (application) on the target, e.g. a vehicle. That 'pillar-based' architecture is shown in Figure 1:

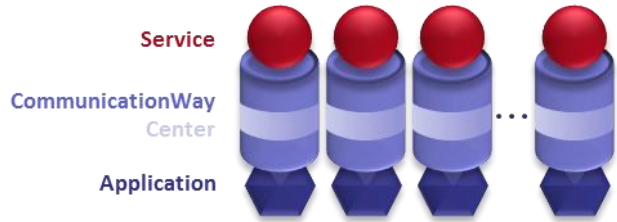

Figure 1: The current 'pillar-based' architecture

This architecture has a couple of shortcomings: The introduction of new services is complex as all the several components need to be recombined or even to be implemented again. This is also true for components which are used by almost every service like authorization, authentication and accounting (AAA) or communication path specific mechanisms like resending of lost information. Therefore, developing the whole pillar for each service again requires valuable resources (e.g. cost and time), as there is no easy, defined way of reusing existing solutions.

Furthermore as those services tend to be isolated from each other, potential synergies are hard to use, as two services might use totally different paradigms, data models and communication technologies. As can be seen in Figure 2, several services tend to be isolated from each other, forming the aforementioned pillars. New services cannot connect easily into such a pillar, based on the non-open design and architecture of the current solutions.

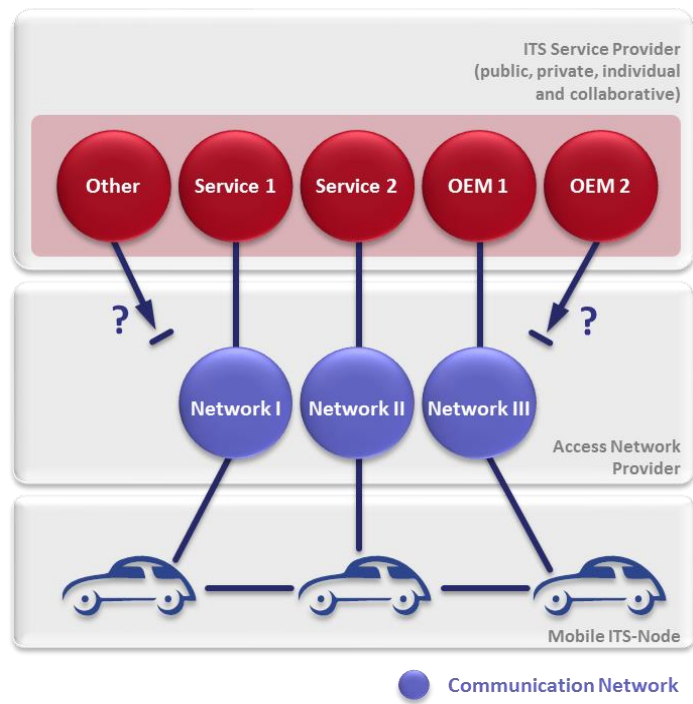

Figure 2: Vertical, isolated structures.

**Future situation: Goals of CONVERGE**

It is expected, that the future European ITS environment will combine a lot of different stakeholders from different fields of work, like vehicle manufactures, road authorities and other companies like vehicle suppliers, network operators or navigation system manufacturers. Those will cooperate to provide various ITS services to different types of devices, like integrated vehicle platforms or personal mobile devices.

To support this future-oriented development, infrastructure-side architecture has to be provided, that offers the necessary framework so that services can be deployed easily. To facilitate the day-one services as well as further and more complex services in the future, the architecture has to be scalable. In addition the architecture should not be limited to one country or one region, but has to enable a trans-regional and international deployment of ITS services.

CONVERGE aims to design this architecture and has planned to introduce the following technological innovation in contrast to the current 'pillar-based' ITS service architecture. At first the architecture should be operator independent, so that the whole business is not dependent on one company or provider. For the service providers and traffic management centers, which want to provide a service, a simple and easy to use integration process has to be established. The adding and removal of services has to be based on an open and standardized interface definition to be as flexible as possible. A transparent use of different complementary access/radio technologies should enable the service users and providers to communicate with each other in such a way that they do not have to concern themselves with the underlying communication technologies. One very important aspect is also a defined and guaranteed information quality especially for information exchanged between different services which is necessary to provide the service quality which is expected in the ITS environment. In an open systems network not only the information quality is relevant, but the quality of given information needs to be verifiable.

These innovations lead to the introduction of the Car2X Systems Network.

**Approach**

As CONVERGE will address the problems described above and therefore tries to create a Car2X Systems Network with the following characteristics:

- **Open**: accessible for all which fulfill a specific set of criteria.
- **Distributed:** The components of the Car2X Systems Network are not placed in one location and are independent from each other.
- **Trans-regional connected**: The systems should not be dependent on regional characteristics and must be usable in all regions as well as communicated information between different regions or countries.
- **Provider independent:** The Car2X Systems Network should not depend on a single provider to operate the network.
- **Scalable:** The Car2X Systems Network must be able to handle a few hundred as well as many thousands of services.
- **Hybrid communicating**: The Car2X Systems Network must work with different communication technologies (e.g. ETSI ITS G5, ETSI ITS IMT Public) to connect service users and providers.
- **Flexible**: The Car2X Systems Network must support different business models, so that new service providers and users choose to join the Car2X Systems Network.

- **Secure**: The Car2X Systems Network should ensure the safety, security and the privacy of the users, the service providers and the communication between them.

Its architecture will break up the current fixed connections between access networks and service providers by specifying standardized, uniform access points between communication networks and service providers. This will allow for an easy addition of new service providers or communication networks as well as a trans-regional, international expansion (see Figure 3).

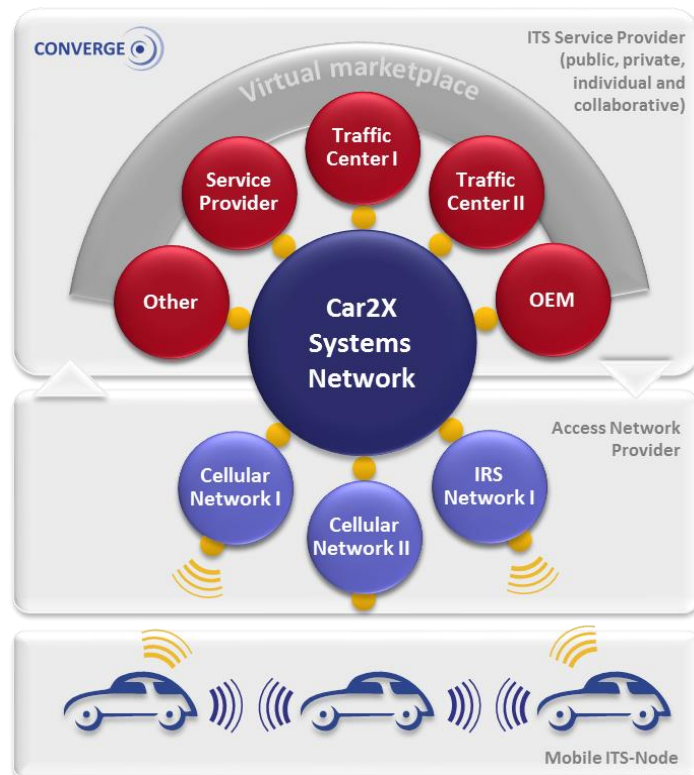

Figure 3: The Car2X Systems Network

For ITS related data, the German Federal Ministry of Transport, Building and Urban Development (BMVBS) has funded a virtual marketplace called "MDM: Mobilitätsdatenmarktplatz" [8] (MDM: Mobility data marketplace) which has created a solution for different participants to trade transport related data in a defined, nationwide manner. Nevertheless this is a solution for traffic related data on a business-to-business base only, which does not help in providing this data to Vehicle ITS Substations (VIS) or Roadside ITS Substations (IRS)[9], nor does it address the problem of communicating between ITS services themself. Therefor CONVERGE will also create a solution which will allow the exchange of quality-marked data between service providers in addition to data exchange between service providers and mobile nodes.

Due to the safety and privacy character of the exchanged information from ITS services, security considerations are an integral part of the CONVERGE architecture. As security may not be guaranteed in an environment where several heterogeneous networks have been connected, an end-to-end security concept needs to be defined in advance. Therefore privacy and security mechanisms will be evaluated and defined for all components of the Car2X Systems Network.

Current research has shown that an introduction of C2X technologies into the market is necessary to allow further progress in the development of ITS services. However several market barriers [10] exist, which prevent C2X technologies from an easy introduction into the market. Those are, beside others:

- Missing user acceptance for day-one services
- Market and government failures, especially for provider-based solutions
- Downward spiral of operation companies

This research has revealed that it will not be possible yet alone profitable for a single provider, or a group of companies forming a single provider, to overcome those market barriers. A possible solution for this dilemma is the introduction of institutional role models for organizational structures instead of providers. Those role models use roles and actors to describe interaction between organizations [11,12,13]. Thereby a role describes a set of actions and the conditions, under which these have to be executed, as well as obligations, liabilities, etc. for organizational roles. Those roles are impartible by definition. A role can be fulfilled by multiple actors, which may be companies, institutions or other organizational entities. Distribution obligations, liabilities and specific behavior to those roles ensure the capability to dynamically react to changes, as well as to introduce flexible interactions between businesses. Additionally, the institutional role model provides transparency between the various organizations and reduces the conflict of goals between different groups.

According to current developments in information and communication technologies (ICT)[14], role models can be used on a technical level to decouple specified behavior of a component from the implementation. It is therefore possible, to define and implement a system which relays on specific functionalities which are not bound to a specific component but to a behavior. This is similar to the structure of the Internet where a lot of distributed servers implement the same behavior in such a way, that the failure of one of those does not interrupt the whole system (e.g. the Domain Name System [15]). Therefore, a provider independent architecture is implemented by CONVERGE.

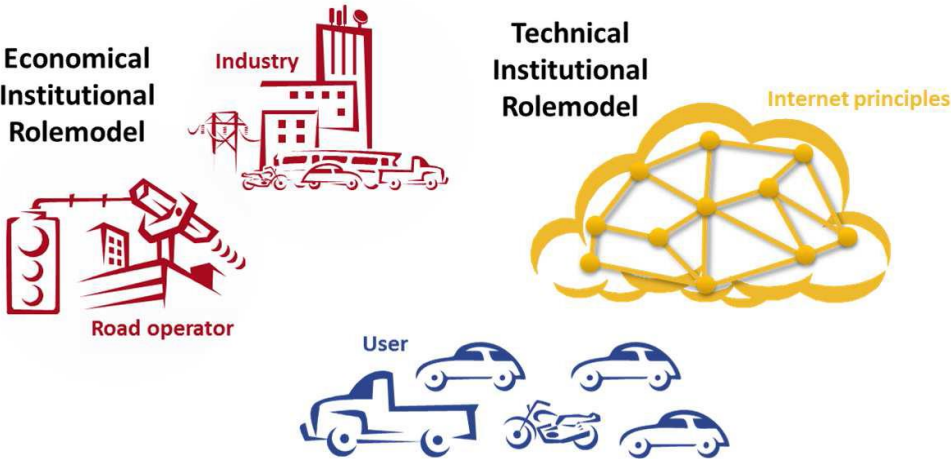

Figure 4: Different role models and actors.

Those considerations, as shown in Figure 4, lead in combination to the CONVERGE concept architecture, as shown in Figure 5. The distributed, technical role models are the foundation for the architecture of the Car2X Systems Network, in the figure represented by golden circles.

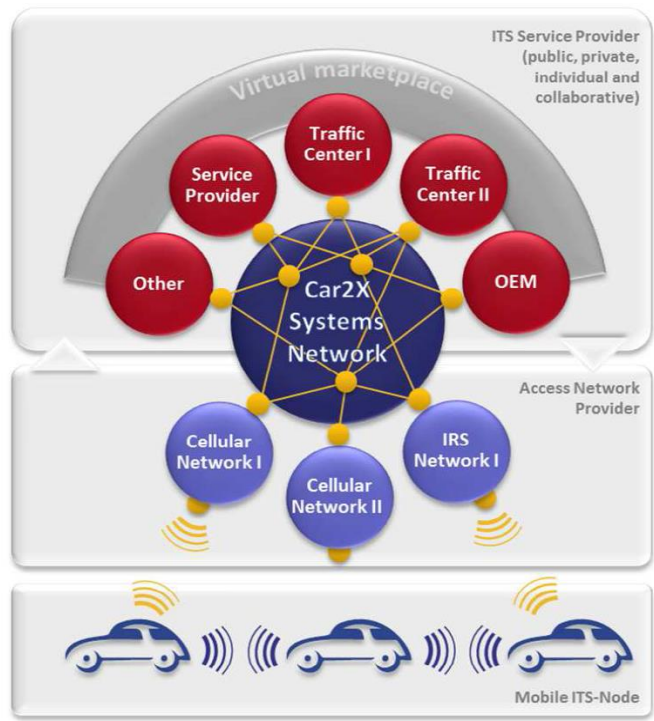

Figure 5: The CONVERGE concept architecture.

CONVERGE will specify software concepts and processes to integrate backend ITS services and communication networks into the Car2X Systems Network, as well as concepts for data management and Quality of Service (QoS) inside the systems network. CONVERGE will also specify mechanisms for the real time capable communication flow between central services and vehicle stations via cellular networks as well as via IRS networks.

To support the introduction of new ITS services, and to prevent them from solving the same typical problems again and again as well as to make service development faster and cheaper, CONVERGE will provide a 'toolbox'. This 'toolbox' will contain solutions for typical problems like:

- Communication
- Standardized access to transmission systems
- Accounting, Authentication & Authorisation (AAA)
- IT-Security and Privacy
- Georouting, which allows sending of information to geographical addressed destinations.
- Multi-path communication with QoS adherence

This 'toolbox' allows easier development with reduced complexity and more features, hopefully leading to more and better services.

The role models, the 'toolbox' and the general considerations of architecture will be combined in a reference architecture for the Car2X Systems Network. This architecture will contain a definition of the basic network architecture, elaborated IT-security and privacy and data management concepts. Furthermore a design for controlling and realizing the flow of information from mobile nodes over ETSI ITS G5 or cellular networks to service providers and vice versa, will be created. For the mobile nodes and the service providers, standardized interfaces will be designed, to be used by the different service implementations. Additionally, a process for the accreditation of new participants will be proposed.

To prove the created architecture, interfaces and mechanisms, CONVERGE will implement a prototype consisting of the key parts of the specified concepts. This prototype will not only include software on the mobile node or the service provider, but also logic in the IRS or cellular networks as well as additionally entities between those and the service provider. A major responsibility of the prototype is the demonstration of the security concepts.

Subsequently, CONVERGE will determine the efficiency and effectiveness of the chosen architecture. Therefore several criteria for assessing the realization of the concepts will be defined. The prototype will also be used to measure implementation related criteria, whereas an expert rating is performed to evaluate the criteria which cannot be measured with the prototype.

**Conclusion**

The Car2X Systems Network is a dynamical extendable association for cooperative systems in ITS, comparable with the internet with open standards and interoperability, which picks up current innovations in the field of ICT and thus provides novel approaches for system- and software-design. CONVERGE aims to develop the architecture of cooperative systems for the mobility of the future.

**Acknowledgements**

CONVERGE [16] is supported by the Federal Ministry of Education and Research and the Federal Ministry of Economics and Technology on the basis of a decision by the German Bundestag.
The CONVERGE consortium consists of the following partners: BMW Research and Technology GmbH, Adam Opel AG, Volkswagen AG, Robert Bosch GmbH, PTV Planung Transport Verkehr AG, Ericsson GmbH, Vodafone GmbH, Hessen Mobile Road and Traffic Management, Road administration of the city of Frankfurt/Main, University of Applied Sciences (htw saar), Federal Highway Research Institute (BASt), Fraunhofer Research Institution AISEC, Fraunhofer-Institute FOKUS and the Federal Network Agency (BNetzA).